# ACCURATE AND ROBUST ALIGNMENT OF VARIABLE-STAINED HISTOLOGIC IMAGES USING A GENERAL-PURPOSE GREEDY DIFFEOMORPHIC REGISTRATION TOOL

*Ludovic Venet[1,2], Sarthak Pati[1,2], Paul Yushkevich[1,2,3,\*], Spyridon Bakas[1,2,4,\*,†]*

1. Center for Biomedical Image Computing and Analytics (CBICA); 2. Department of Radiology; 3. Penn Image Computing and Science Lab (PICSL); 4. Department of Pathology and Laboratory Medicine, Perelman School of Medicine, University of Pennsylvania, Philadelphia, PA, USA.

\* Equally contributing senior authors.
† Corresponding author: sbakas@upenn.edu

## ABSTRACT
Variously stained histology slices are routinely used by pathologists to assess extracted tissue samples from various anatomical sites and determine the presence or extent of a disease. Evaluation of sequential slides is expected to enable a better understanding of the spatial arrangement and growth patterns of cells and vessels. In this paper we present a practical two-step approach based on diffeomorphic registration to align digitized sequential histopathology stained slides to each other, starting with an initial affine step followed by the estimation of a detailed deformation field.

## 1. INTRODUCTION

In the area of Digital Pathology an acceptable schema for evaluating the anatomical structure of a given dataset is to associate the visual appearance of consecutive tissue sections. Due to the way tissues can be processed, differentially stained, and other pre-analytical steps, successive slices can suffer from non-linear deformations, as well as dramatic appearance changes. This evaluation schema can be assessed by aligning consecutive slices to a common coordinate system.

## 2. MATERIALS & METHODS

### 2.1. Data

The Automatic Non-rigid Histological Image Registration (ANHIR) challenge [1] describes a publicly available multi-institutional dataset [1-5] and a community benchmark to fairly evaluate and compare various deformable registration methods. A set of high-resolution (up to 40x magnification) whole-slide images of different anatomical sites are made publicly available. These images are organized in sets of consecutive sections of a distinct anatomical site, where each slice was stained with a different dye and any two images within a set can be meaningfully registered (Fig.1).

### 2.2. Pre-processing

We note that the mammary gland slides dyed for ER and c-erbB-2/HER-2-neu use a Diaminobenzidine (DAB) stain, which has a brown-dominating appearance and significant background staining that makes them very distinct from all other stained slides. Therefore, we apply a color deconvolution algorithm [6] to remove the DAB color components and avoid potential misregistrations.

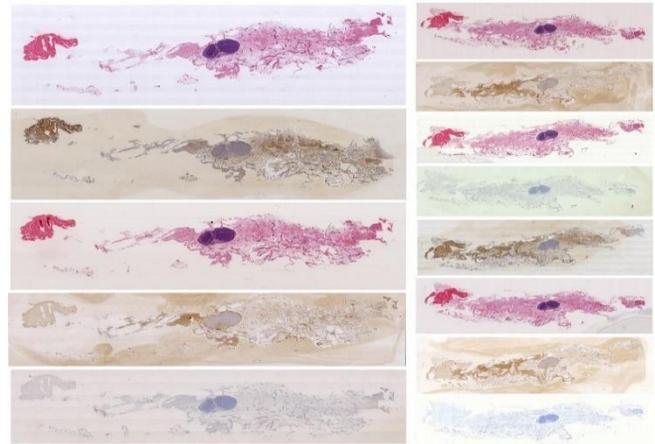

Figure 1. Example sequential stained slides given by the ANHIR challenge. Digitized slides from mammary gland are depicted here. Figure taken from anhir.grand-challenge.org

Taking into consideration the large size of the scaled images used in the evaluation of our approach, we resampled them based on a factor (*f*) of 1/25. To prevent aliasing due to such resampling, we applied a Gaussian kernel before down-sampling. All images were then converted to NIfTI format [7] for all computations.

We applied padding to each stained image to ensure that i) the sizes of paired images are the same and ii) the target tissue is in the image center. Once all image pairs are of the same size, we add an additional padding (4x the size of the similarity metric's kernel - Eq.1) to make sure that the tissue is far enough from the image boundaries, and accommodate appropriate calculation of the deformation field after changes caused by the affine registration. A binary mask excluding the image boundaries (size of the similarity metric kernel - Eq.1) is also used during affine registration. The padded areas are filled with image intensity matching that of the four corners of the unpadded image.

### 2.3. Registration

We register the variously-stained histopathology slices using a general-purpose 2D/3D medical image registration tool "*Greedy*" (github.com/pyushkevich/greedy) [8]. *Greedy* is a C++ implementation of the greedy diffeomorphic registration algorithm [9]. *Greedy* is integrated into the ITK-SNAP (itksnap.org) segmentation software [10, 11], as well as the Cancer Imaging Phenomics Toolkit (CaPTk - www.cbica.upenn.edu/captk) [12, 13].

*Greedy* shares multiple concepts and implementation strategies with the Symmetric Normalization (*SyN*) registration tool in the ANTS package [14, 15], but focuses on computational efficiency by eschewing the symmetric property of SyN and utilizing highly optimized code for computation of image similarity metrics such as Normalized Cross-Correlation (NCC) [16].

For our approach, we use the NCC metric as:

$$\text{NCC Kernel Size} = \lfloor \frac{\text{Size}(I_i)}{S} \rfloor \quad (1)$$

where $S$ is the scale by which the size of the fixed image $I_i$ is scaled so that the NCC kernel can pick up enough information about the images to perform a good registration.

The registration of the image pairs is performed while considering three different scales in a configuration of a multi-resolution pyramid.

Following [17], this paper uses the notation:

$$T^*_{i \to j} = R(I_i \to I_j; \theta) \quad (2)$$

where $(T^*_{i \to j})$ describes the transformation between fixed ($I_i$) and moving ($I_j$) image, and $\theta$ defines the registration parameters yielding transformation $T^*_{i \to j}$. $R$ defines a minimization process such that Eq.2 is unfolded as:

$$T^*_{i \to j} = \underset{T_{i \to j}}{\text{argmin}}\, \mu(I_i, I_j \circ T_{i \to j}) + \lambda \rho(T_{i \to j}) \quad (3)$$

where $\mu$ is the similarity metric, $\lambda$ is a scalar parameter, and $\rho$ is an optional regularization term.

We firstly perform affine registration between the image pairs, using an optimization of the dissimilarity metric based on a Limited-memory Broyden–Fletcher–Goldfarb–Shanno algorithm (LBFGS) [18], denoted by:

$$A_{i \to j} = R_{aff}(I_i \to I_j; \mu, A_0) \quad (4)$$

where $A_0$ is the initial rigid transformation between the images. The initial transformation is obtained using a brute force search, where 5000 random rotations and translations are applied to the moving image and the combination giving the best NCC metric value is used as $A_0$.

Following the affine registration, we applied the diffeomorphic registration of slice $j$ to $i$:

$$\varphi_{i \to j} = R_{diff}(I_i \to I_j; \mu, \sigma_s, \sigma_t, N) \quad (5)$$

Where $\sigma_s$ and $\sigma_t$ are the regularization parameters for the registration and $N$ is the number of iterations required at each multi-resolution pyramid (for instance, $N=\{100,50,10\}$ means 100 iterations at 4x, 50 at 2x and 10 at full resolution).

Furthermore, *Greedy* uses an optimized smoothing of the deformation fields based on the ITK recursive Gaussian smoothing classes [19]. The actual registration is computed in an iterative manner using the update equations:

$$\psi^\gamma = Id + \varepsilon^\gamma \cdot \left[ G_{\sigma_S} * D_{\varphi^T_{i \to j}} \mu(I_i, I_j \circ \varphi^\gamma_{i \to j}) \right] \quad (6)$$

$$\varphi^{\gamma+1}_{i \to j} = G_{\sigma_t} * (\varphi^\gamma_{i \to j} \circ \psi^\gamma) \quad (7)$$

$$\varphi^0_{i \to j} = Id \quad (8)$$

where $\gamma$ is the current iteration, $D_{\varphi^T_{i \to j}}$ is the gradient of the metric in comparison to $\varphi$, $\varepsilon^\gamma$ is the step size, $G_{\sigma_t} * \varphi$ denotes the convolution of $\varphi$ with an isotropic Gaussian kernel with a standard deviation of $\sigma$ and $Id$ is the identity transformation. For sufficiently smaller $\varepsilon^\gamma$ and larger $\sigma_s$ values, $\psi^\gamma$ is smooth and has positive Jacobian determinant for all $x \in \Omega_i$, thereby making the registration diffeomorphic in nature. As diffeomorphisms form a group under composition, $\varphi^{T+1}_{i \to j}$ is also diffeomorphic in nature [9].

## 2.4. Evaluation

The quantitative performance evaluation framework reported here was based on the average of the median relative Target Registration Error (*rTRE*). Specifically, *TRE* is defined as:

$$TRE = d_e(x^T_l, x^W_l) \quad (9)$$

where $x^T$ and $x^W$ are the coordinates of the landmarks in the target and warped image and $d_e(.)$ defines the Euclidean distance. All TRE are then normalized by the diagonal of the image to define the *rTRE:*

$$rTRE = \frac{TRE}{\sqrt{w^2 + h^2}} \quad (10)$$

where w and h denote the image weight and height, respectively.

## 3. RESULTS

We performed a grid search for $\sigma_s$ and $\sigma_t$ in the range of [20,20] and we found that the optimal values based on the training data were 6 and 5 pixels, respectively.

The average across all pairs of images of the median of the rTRE for the affine and the deformable registration is equal to 0.00473 and 0.00279, respectively. Notably, our rTRE score of 0.00279 is the highest of the challenge's leaderboard.

We further note that the robustness of our method as defined by the challenge was equal to 1.

## 4. CONCLUSION

A surprising finding is that a general-purpose tool originally developed for registration of 3D medical images such as MRI, achieved excellent performance in the domain of histology registration. *Greedy* has previously been used for 3D reconstruction of histology and histology-MRI matching [17], and relatively little parameter tuning was needed to adopt it to this challenge. Image processing scripts will be made available through our GitHub page at github.com/CBICA/HistoReg.

Future work includes the further performance evaluation of the *Greedy* algorithm and its comparison with alternative approaches, e.g., based on detection of salient points [20]. The overarching goal of this work is to reconstruct the 3D anatomical tissue structure from 2D histology slices [17, 21, 22] irrespective of the staining applied to them, in order to give more context and evaluate the association of anatomical structures in the microscopic scale with the molecular characterization of the associate tissue samples. Notably, this is of interest in cancer, where such associations are already evaluated in the macroscopic scale based on radiographic representations [23-26].

## 5. ACKNOWLEDGEMENTS

This work was supported by: U24 CA189523, R01 EB017255, R01 AG056014, and P30 AG010124.

# 6. REFERENCES

[1] J. Borovec, A. Munoz-Barrutia, and J. Kybic, "Benchmarking of Image Registration Methods for Differently Stained Histological Slides," in *2018 25th IEEE International Conference on Image Processing (ICIP)*, 2018, pp. 3368-3372.

[2] R. Fernandez-Gonzalez, A. Jones, E. Garcia-Rodriguez, P. Y. Chen, A. Idica, S. J. Lockett*, et al.*, "System for combined three-dimensional morphological and molecular analysis of thick tissue specimens," *Microscopy Research and Technique,* vol. 59, pp. 522-530, 2002/12/15 2002.

[3] L. Gupta, B. M. Klinkhammer, P. Boor, D. Merhof, and M. Gadermayr, "Stain independent segmentation of whole slide images: A case study in renal histology," in *2018 IEEE 15th International Symposium on Biomedical Imaging (ISBI 2018)*, 2018, pp. 1360-1364.

[4] I. Mikhailov, N. Danilova, and P. Malkov, "The immune microenvironment of various histological types of EBV-associated gastric cancer," in *VIRCHOWS ARCHIV*, 2018, pp. S168-S168.

[5] G. Bueno and O. Deniz. *AIDPATH: Academia and Industry Collaboration for Digital Pathology,* http://aidpath.eu/?page_id=279.

[6] A. C. Ruifrok and D. A. Johnston, "Quantification of histochemical staining by color deconvolution," *Analytical and quantitative cytology and histology,* vol. 23, pp. 291-299, 2001.

[7] R. W. Cox, J. Ashburner, H. Breman, K. Fissell, C. Haselgrove, C. J. Holmes*, et al.*, *A (sort of) new image data format standard: NiFTI-1* vol. 22, 2004.

[8] P. A. Yushkevich, J. Pluta, H. Wang, L. E. M. Wisse, S. Das, and D. Wolk, "Fast automatic segmentation of hippocampal subfields and medial temporal lobe subregions in 3 Tesle and 7 Tesla T2-weighted MRI," *Alzheimer's & Dementia: The Journal of the Alzheimer's Association,* vol. 12, pp. P126-P127, 2016.

[9] S. Joshi, B. Davis, M. Jomier, and G. Gerig, "Unbiased diffeomorphic atlas construction for computational anatomy," *NeuroImage,* vol. 23, pp. S151-S160, 2004/01/01/ 2004.

[10] P. A. Yushkevich, J. Piven, H. C. Hazlett, R. G. Smith, S. Ho, J. C. Gee*, et al.*, "User-guided 3D active contour segmentation of anatomical structures: Significantly improved efficiency and reliability," *NeuroImage,* vol. 31, pp. 1116-1128, 2006/07/01/ 2006.

[11] P. A. Yushkevich, A. Pashchinskiy, I. Oguz, S. Mohan, J. E. Schmitt, J. M. Stein*, et al.*, "User-Guided Segmentation of Multi-modality Medical Imaging Datasets with ITK-SNAP," *Neuroinformatics,* vol. 17, pp. 83-102, 2019/01/01 2019.

[12] C. Davatzikos, S. Rathore, S. Bakas, S. Pati, M. Bergman, R. Kalarot*, et al.*, "Cancer imaging phenomics toolkit: quantitative imaging analytics for precision diagnostics and predictive modeling of clinical outcome," *Journal of Medical Imaging,* vol. 5, p. 011018, 2018.

[13] S. Rathore, S. Bakas, S. Pati, H. Akbari, R. Kalarot, P. Sridharan*, et al.*, "Brain Cancer Imaging Phenomics Toolkit (brain-CaPTk): An Interactive Platform for Quantitative Analysis of Glioblastoma," *Brainlesion : glioma, multiple sclerosis, stroke and traumatic brain injuries : third International Workshop, BrainLes 2017, held in conjunction with MICCAI 2017, Quebec City, QC, Canada, September 14, 2017, Revised selected papers. Bra...* vol. 10670, pp. 133-145, 2018.

[14] B. B. Avants, C. L. Epstein, M. Grossman, and J. C. Gee, "Symmetric diffeomorphic image registration with cross-correlation: Evaluating automated labeling of elderly and neurodegenerative brain," *Medical Image Analysis,* vol. 12, pp. 26-41, 2008/02/01/ 2008.

[15] B. B. Avants, N. J. Tustison, G. Song, P. A. Cook, A. Klein, and J. C. Gee, "A reproducible evaluation of ANTs similarity metric performance in brain image registration," *NeuroImage,* vol. 54, pp. 2033-2044, 2011/02/01/ 2011.

[16] D.-M. Tsai and C.-T. Lin, "Fast normalized cross correlation for defect detection," *Pattern Recognition Letters,* vol. 24, pp. 2625-2631, 2003/11/01/ 2003.

[17] D. H. Adler, L. E. M. Wisse, R. Ittyerah, J. B. Pluta, S.-L. Ding, L. Xie*, et al.*, "Characterizing the human hippocampus in aging and Alzheimer's disease using a computational atlas derived from ex vivo MRI and histology," *Proceedings of the National Academy of Sciences,* vol. 115, pp. 4252-4257, 2018.

[18] A. Mokhtari and A. Ribeiro, "Global convergence of online limited memory BFGS," *The Journal of Machine Learning Research,* vol. 16, pp. 3151-3181, 2015.

[19] R. Deriche, "Fast algorithms for low-level vision," *IEEE Transactions on Pattern Analysis and Machine Intelligence,* vol. 12, pp. 78-87, 1990.

[20] S. Bakas, M. Doulgerakis-Kontoudis, G. J. A. Hunter, P. S. Sidhu, D. Makris, and K. Chatzimichail, "Evaluation of indirect methods for motion compensation in 2D focal liver lesion Contrast-Enhanced Ultrasound (CEUS) imaging," *Ultrasound in Medicine & Biology,* vol. In Press, 2019.

[21] P. A. Yushkevich, B. B. Avants, L. Ng, M. Hawrylycz, P. D. Burstein, H. Zhang*, et al.*, "3D Mouse Brain Reconstruction from Histology Using a Coarse-to-Fine Approach," in *Biomedical Image Registration*, Berlin, Heidelberg, 2006, pp. 230-237.

[22] D. H. Adler, J. Pluta, S. Kadivar, C. Craige, J. C. Gee, B. B. Avants*, et al.*, "Histology-derived volumetric annotation of the human hippocampal subfields in postmortem MRI," *NeuroImage,* vol. 84, pp. 505-523, 2014/01/01/ 2014.

[23] S. Bakas, H. Akbari, J. Pisapia, M. Martinez-Lage, M. Rozycki, S. Rathore*, et al.*, "In vivo detection of EGFRvIII in glioblastoma via perfusion magnetic resonance imaging signature consistent with deep peritumoral infiltration: the φ-index," *Clinical Cancer Research,* vol. 23, pp. 4724-4734, 2017.

[24] H. Akbari, S. Bakas, J. M. Pisapia, M. P. Nasrallah, M. Rozycki, M. Martinez-Lage*, et al.*, "In vivo evaluation of EGFRvIII mutation in primary glioblastoma patients via complex multiparametric MRI signature," *Neuro-Oncology,* vol. 20, pp. 1068-1079, 2018.

[25] Z. A. Binder, A. H. Thorne, S. Bakas, E. P. Wileyto, M. Bilello, H. Akbari*, et al.*, "Epidermal Growth Factor Receptor Extracellular Domain Mutations in Glioblastoma Present Opportunities for Clinical Imaging and Therapeutic Development," *Cancer Cell,* vol. 34, pp. 163-177.e7, 2018/07/09/ 2018.

[26] S. S. M. Elsheikh, S. Bakas, N. J. Mulder, E. R. Chimusa, C. Davatzikos, and A. Crimi, "Multi-stage Association Analysis of Glioblastoma Gene Expressions with Texture and Spatial Patterns," in *Brainlesion: Glioma, Multiple Sclerosis, Stroke and Traumatic Brain Injuries*, Cham, 2019, pp. 239-250.